\documentclass[prl,twocolumn,showpacs,preprintnumbers,amsmath,amssymb]{revtex4}
\usepackage{graphicx}
\usepackage{dcolumn}
\usepackage{bm}
\usepackage{amssymb}
\usepackage[center]{subfigure}
\usepackage{xspace}

\newcommand\etal{\textit{et al.}\xspace}
\newcommand\ie{{i.e.\ }}

\renewcommand{\deg}{^{\circ}}
\newcommand\Si{\ensuremath{\mathcal{S}}}

\begin{document}
\title{Meandering rivulets on a plane\,: a simple balance between inertia and capillarity\,?}
\author{Nolwenn \surname{LE GRAND-PITEIRA}}
\email{nlegrand@pmmh.espci.fr}
\author{Adrian DAERR}
\author{Laurent LIMAT}
\affiliation{ Laboratoire de Physique et M\'ecanique des Milieux
H\'et\'erog\`enes, 10 rue Vauquelin 75005 Paris France, UMR CNRS 7636\\ and Mati\`ere et Syst\`emes Complexes, University Paris 7, UMR CNRS
7057}

\date{\today}

\begin{abstract}
Experiments on streams of water flowing down a rigid substrate have
been performed for various plate inclinations and flow rates, and we
focused on the regime of stationary meanders. The outcome is that (i)
the flow is highly hysteretic\,: the shape of the meanders varies with
flow rate only for increasing flow rates, and the straight rivulet
regime does not appear for decreasing flow rate. (ii) A simple force
balance, including inertia, capillary forces, and also hysteresis of
wetting, accounts well for the experimental instability threshold flow
rate and for the final radius of curvature of the meanders.
\end{abstract}

\maketitle

Meandering is ubiquitous in nature and most familiar from rivers, but
can also be seen at a much smaller scale like on window-panes during a
rainfall. Recently, Drenckhan et al. \cite{Drenckhan} identified
regular meandering patterns of rivulets, confined between two vertical
plates, and charged with surfactants, mimicking undulations of foam
Plateau borders. This observation confirms the generality of the
pattern and suggests that a well defined mechanism is at work in
rivulets, able to select a given meandering length
scale. Understanding this mechanism and the properties of the
resulting meanders is by itself a fundamental challenge which can also
have implications in other fields of physics. For instance, Bruinsma
\cite{Bruinsma} pointed out a possible analogy with the statistical
mechanics for directed polymers inside a random matrix.  The stability
and characteristics of meandering rivulets may also be of importance
to several industrial processes.  In heat exchangers, changes between
different flow regimes can cause drastic modifications of heat
transfer \cite{Ganic}. Meander formation can also be an undesirable
feature in coating processes \cite{Kistler}.

For rivers \cite{Leopold, Liverpool}, it seems asserted that erosion
is the key mechanism, but the meandering on non-erodible surfaces is
still an essentially open problem. Culkin \cite{Culkin} and Nakagawa
\cite{Nakagawa84, Nakagawa92} reported significant experimental work
on streams running down an inclined plate. Nakagawa identified four
regimes depending on flow rate\,: drops, meanders, unstable stream
(the main rivulet oscillates and splits into several smaller ones) and
a restable stream (the rivulet restabilizes into a straight rivulet of
variable width, forming a braided pattern \cite{Mertens}). He stresses
that he never obtained straight rivulets (of constant width), though
this regime has been reported by Schmuki and Laso \cite{Schmuki} who
also investigated the effects of viscosity and surface tension,
showing that meandering is suppressed at high viscosities. All in all,
meandering has been observed in a wide variety of configurations, but
little or no general conclusions have been drawn. In several studies the
order parameter is the ratio of the length of the meanders over the
length of the inclined plane (termed sinuosity, well kown in a
geological context), a quantity difficult to interpret physically.

Stability analyses of rivulets, some neglecting longitudinal flow
\cite{Davis, Young}, are mainly focused on varicose modes and not
sinuous modes, \ie are not dealing with the meandering
instability. Only two recent papers deal with meandering threshold
\cite{Bruinsma, Kim}.

This paper studies the shape and behavior of meandering rivulets as a
function of two control parameters (flow rate and plate inclination),
and discusses physical interpretations, focusing on the role of
hysteresis. Up to now, there has been no quantitative study of the
morphology of meanders on non-erodible surfaces in well controlled and
reproducible conditions, and in addition, no comparison with simple
hydrodynamic models. Our paper is a first step in this direction. In
the present paper, experiments were only performed at low viscosity
(that of water), corresponding to common natural situations.
\smallskip

\begin{figure}[hb!]%
		\begin{center}%
			a) \includegraphics[width=0.22\textwidth]{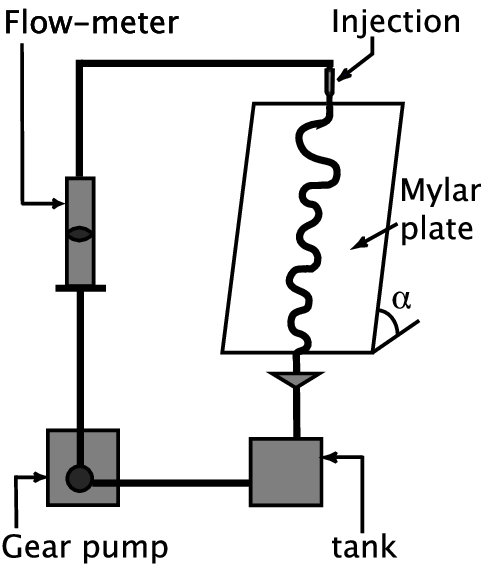}%
			b)\includegraphics[width=0.2\textwidth]{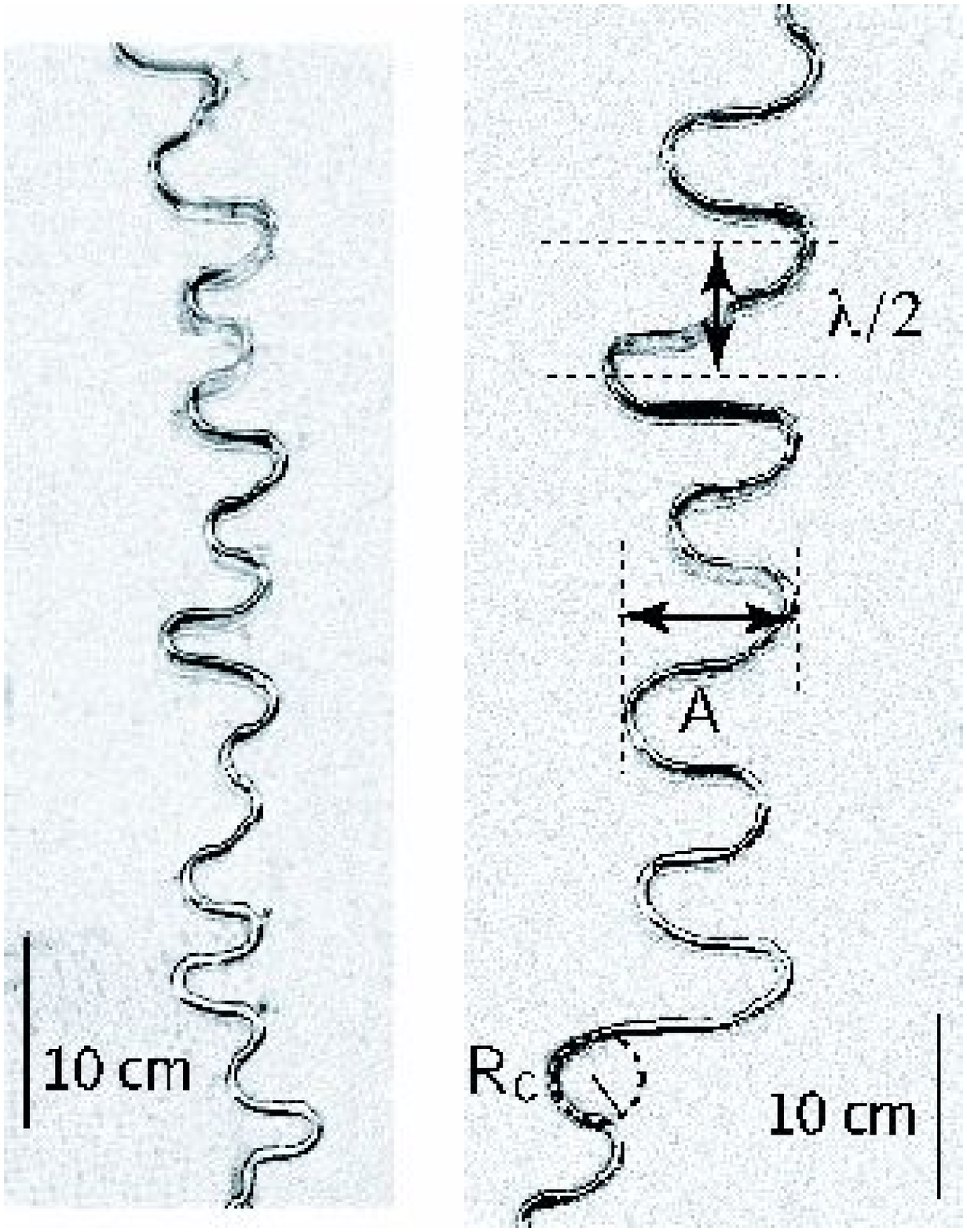}%
			\caption{a) Experimental setup  b)\,Stationary meanders for $\alpha=32\deg$. Q=1.08\,mL/s for the left picture and Q=1.40\,mL/s for the right one.}%
		        \label{Montage}%
		\end{center}%
	\end{figure}

\nopagebreak
{\it Experimental set-up~--} Figure \ref{Montage}\textit{a} shows a
schematic diagram of the experimental set-up. De-ionized water is
injected at the top of an inclined plate ($1.20$\,m long and $50$\,cm
wide).  The substrate is a Mylar sheet (plastic sheets of polyethylene
terephthalate (PET)), flattened
on a rigid plate. It insures partial wetting conditions
\addtolength\abovecaptionskip{-1ex}%
\addtolength\belowcaptionskip{-4ex}%
\addtolength\floatsep{-1ex}%
\addtolength\textfloatsep{-1ex}%
 \pagebreak for
water (advancing and receding contact angles of respectively
$\theta_{a}=70\deg$ and $\theta_{r}=35\deg$) and reduces problems of
static electricity compared to other common plastics. The tilt angle
$\alpha$ of the plate can be changed at will between $0\deg$ and
$87\deg$. The water is collected in a tank and pumped back to the top
of the plate by a gear pump providing an adjustable and constant flow
rate $Q$, checked with a precision flow-meter. Pictures and movies of
the experiments were taken by a digital camera placed $1\,$m above the
plate and perpendicular to the latter.
\smallskip


{\it Meandering thresholds~--} For increasing flow rates, the
following regimes are observed. (i) Drops. Individual drops
periodically detach from the injector \cite{LeGrand}.  (ii) Straight
rivulets. The liquid flows down forming a straight continuous ridge
along the direction of steepest descent. (iii) Meandering
rivulets. Above a critical flow rate $Q_{c1}$, depending on the plate
inclination, the straight stream is unstable. Perturbations (surface
defects, injection noise, air movement, \dots) appear as small bends
of typical size comparable to the rivulet width $w$. These bends
initially amplify laterally and downwards and eventually reach a
stationary shape. It can take $10$ minutes to one hour for a meander
to fully develop along the whole length of the plate, but the
resulting path is completely stationary
(Fig.~\ref{Montage}\textit{b}). Snapshots of a settled meander, taken
every 2 minutes during 24 hours, confirmed that it did not move at all
throughout that time lapse. Using either a steady pump or a constant
level tank, differently designed injectors and a long plate to check
whether the behavior of the rivulet depended on the distance to the
injector, we have verified that the meandering regime existed
independently of injection conditions, confirming previous findings by
other groups. 
To obtain yet other injection conditions, it might be interesting to
study the collision/merging of two rivulets.
 (iv) Dynamic regime. Above a second critical flow rate
$Q_{c2}$, meanders no longer remain stable. The rivulet sweeps from
side to side, similar to the free end of a garden hose
\cite{Kuronuma}, frequently breaking up into sub-rivulets. (v)
Restable regime. For even higher flow rates, the rivulet restabilizes
and becomes straight again, but its width now varies like braids
\cite{Mertens}.

\begin{figure}[ht]%
		\begin{center}%
			a)\includegraphics[width=0.3\textwidth]{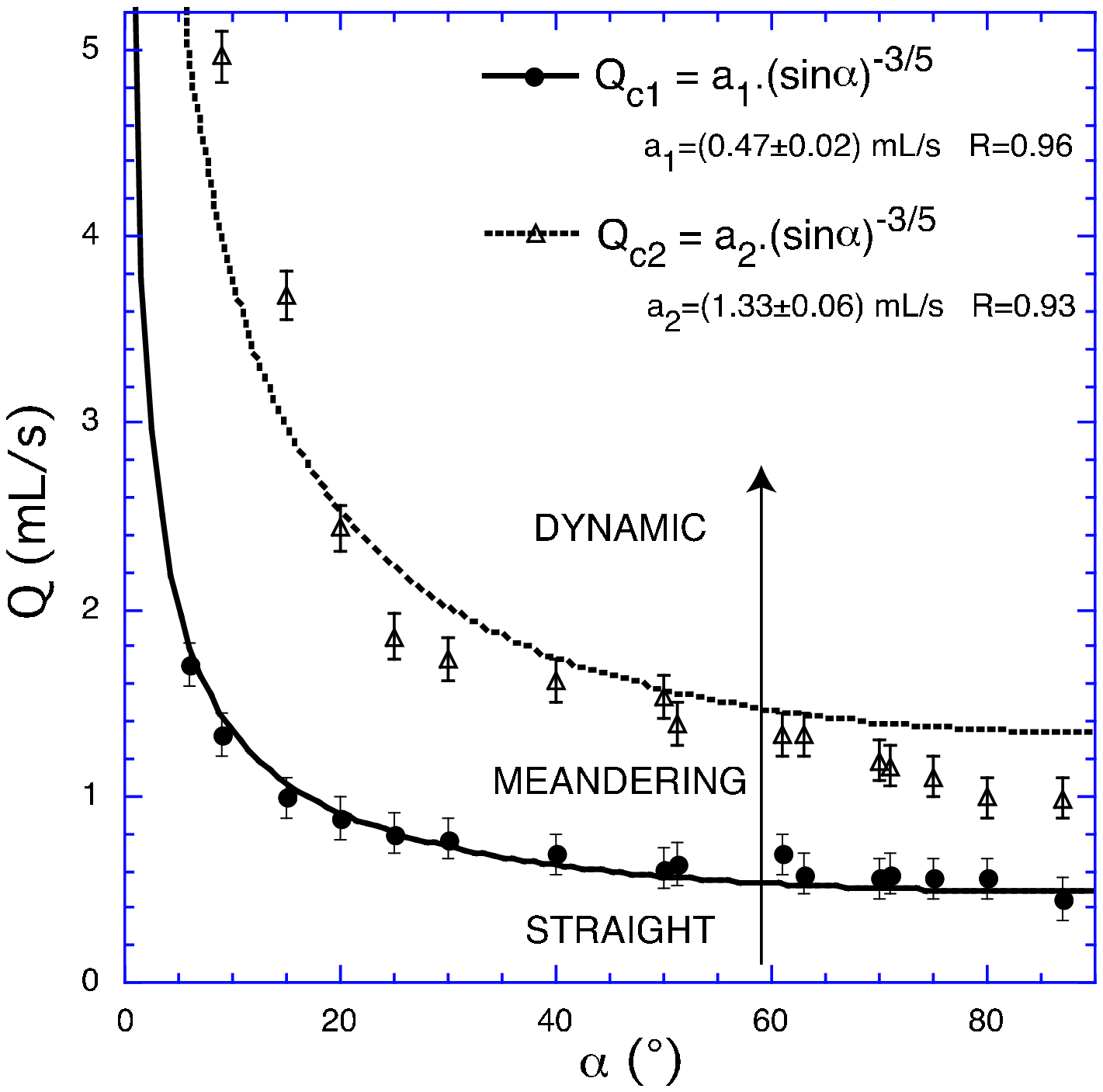}%
			b)\includegraphics[width=0.13\textwidth]{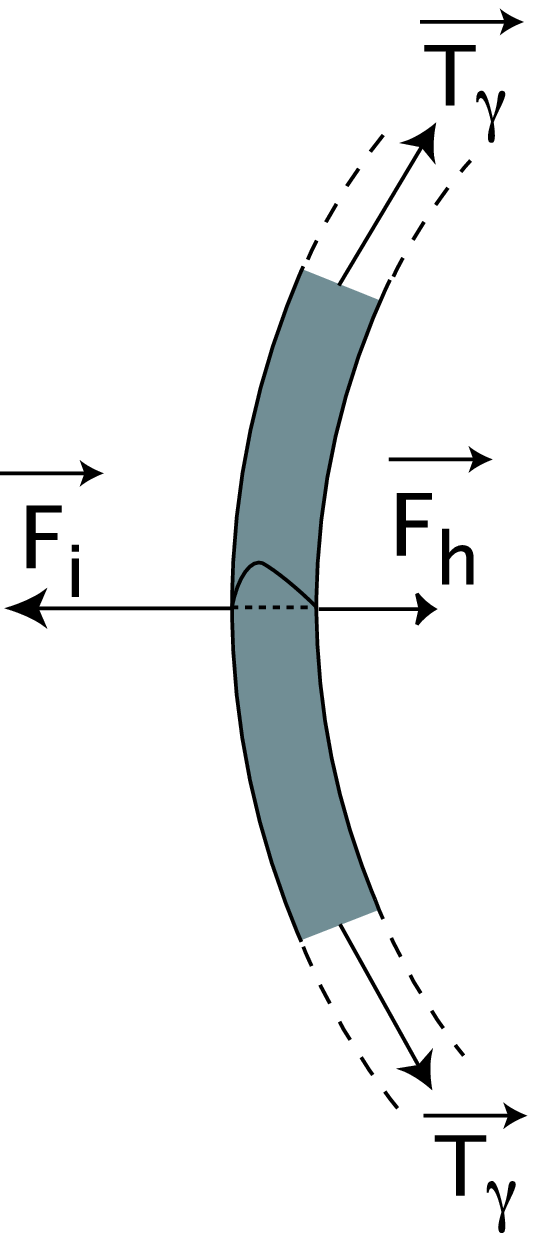}%
			\caption{ a)\,Meandering thresholds for increasing flow rates   b)\,Forces acting on a meander, in the plane of the plate}%
			\label{Seuils}%
		\end{center}%
	\end{figure}

Fig.~\ref{Seuils}\textit{a} displays the critical flow rates for the
onset and disappearance of stationary meanders as a function of plate
inclination $\alpha$. The decrease of the second critical flow rate
$Q_{c2}$ with $\alpha$ is similar to that of $Q_{c1}$, but we have no
explanation for it yet. The dependency of $Q_{c1}$ on $\alpha$ can be
understood from the balance of forces acting on the rivulet\,:
gravity, surface tension, inertia and contact line pinning forces. For
the lateral stability of a straight rivulet, gravity does not
intervene (Fig.~\ref{Seuils}\textit{b}). Surface tension opposes the
bending of the rivulet and, once integrated across the cross-section,
can be seen as a line tension of the liquid rim~\cite{linetension}
creating a normal force $F_{\gamma}$ straightening the rivulet. Taking
into account the interfacial energies and the capillary pressure
inside the rivulet, one gets $F_{\gamma}=C(\theta) \gamma w /r_{c}$,
where $C(\theta)$ is a constant ($C(\theta) \approx \theta^2 /3$ in
the limit of a small average contact angle $\theta$), $\gamma$ the
surface tension, $w$ the width of the rivulet and $r_c$ the initial
radius of curvature. Pinning forces are reactive (therefore
stabilizing forces) and act normal to the contact line. Their upper
bound is given by the advancing and receding contact angles: $F_h \leq
F_h^{\mathrm{max}} = \gamma(\cos\theta_r-\cos\theta_a)$. Instability
will arise when inertia ($F_i=\rho S v^{2}/r_c$ where $\rho$ stands
for the density of the liquid, $S$ for the cross-section of the
rivulet and $v$ for the RMS velocity ($\simeq$ average velocity)
inside the rivulet) becomes stronger than both line tension and
pinning ($F_i \geq F_{\gamma} + F_h^{\mathrm{max}}$). The onset of
meandering is therefore given by:

	\begin{equation}
		\rho \frac{Q_{c1}^2}{S r_c} =  \gamma \bigg [ \frac{C(\theta)w}{r_c} + (\cos\theta_r-\cos\theta_a) \bigg]
		\label{eq:instab}
	\end{equation}

\begin{figure*}[t!]%
	\begin{center}%
		\includegraphics[width=\textwidth]{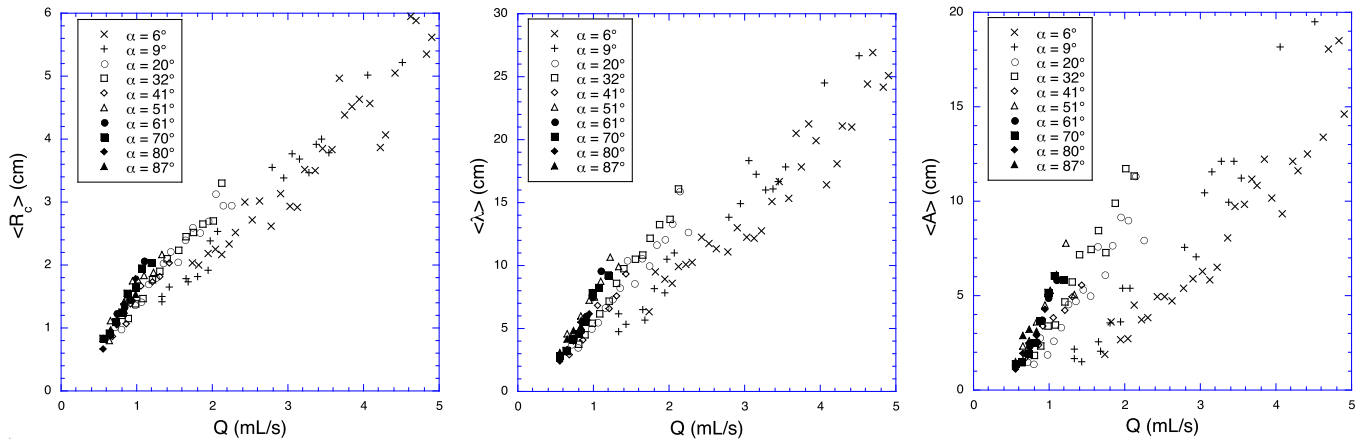}%
		\caption{Experimental data for the mean radius of curvature, wavelength and amplitude for numerous plate inclinations.}%
		\label{AlRc}%
	\end{center}%
\end{figure*}

Without the pinning term, this balance was also suggested by Drenckhan
\etal \cite{Drenckhan} for meanders in foams. If the pinning term can
be neglected, or else if $r_c$ scales as $w$ with flow rate, the first
critical flow rate scales as $\rho Q_{c1}^2 /S \propto \gamma w$. Let
us now assume that the flow inside the rivulet is a Poiseuille flow
($v \propto w^{2} g\sin\alpha/\nu)$, where $\nu$ denotes the kinematic
viscosity), and approximate the cross-section of the rivulet to a disc
segment with $\pi/4$ contact angle ($S=(\pi-2)w^2/8$). Using flow rate
conservation ($Q=Sv$) leads to the scaling
	\begin{equation}
		Q_{c1} \propto \Big[ \left(\gamma/\rho\right)^{4/5} \left(\nu/g\right)^{3/5} \Big]. (\sin\alpha)^{-3/5},
		\label{onsetscaling}
	\end{equation}
This scaling matches the experimental data very well
(Fig~\ref{Seuils}\textit{a}). If we keep the theoretical prefactor
evaluating to about 2.3, we obtain flow rates 5 times too small. Rough
estimates of the force terms in equ~(\ref{eq:instab}), using
experimental data for typical threshold conditions, show that pinning
is not completely negligible~\cite{forceestimates}. We thus expect
equ~(\ref{onsetscaling}) to underestimate the threshold. It is
satisfying that the order of magnitude of the forces, and their
scaling, are correctly estimated. The scaling of
equ~(\ref{onsetscaling}) is close to that predicted by Bruinsma
\cite{Bruinsma}.

If the flow rate is now decreased while in the meandering regime,
meanders remain remarkably stable\,: they keep their path, but become
thinner, until they break up into drops. In particular there is no
sinuous/straight rivulet transition\,: below $Q_{c1}$ the rivulets
meander instead of becoming straight. This strong hysteresis is a
consequence of pinning effects. For all features of a meander with a
radius of curvature greater than
$C(\theta)w/(\cos\theta_{r}-\cos\theta_{a}) \approx 0.4w$, \ie for all
but very small scales, contact line pinning forces (now acting in the
opposite direction) will dominate restoring capillary forces
(equ.~(\ref{eq:instab})), even if inertial forces tend to zero.
\smallskip


{\it Shape of the stationary meanders~--} All the experimental data
presented here were obtained for increasing flow rates. For numerous
meanders achieved for plate inclinations varying between 6$\deg$ and
87$\deg$, we measured the mean radius of curvature $\langle
R_c\rangle$ at the apex of the curves (mean taken over all curves of a
given rivulet), mean wavelength $\langle\lambda\rangle$, and mean
amplitude $\langle A\rangle$ (see figures~\ref{Montage}\textit{b} and
\ref{AlRc}). All three parameters increase monotonically with the
inclination of the plate $\alpha$ and the flow rate $Q$.
		\begin{figure}[ht]%
		\begin{center}%
			\includegraphics[width=0.37\textwidth]{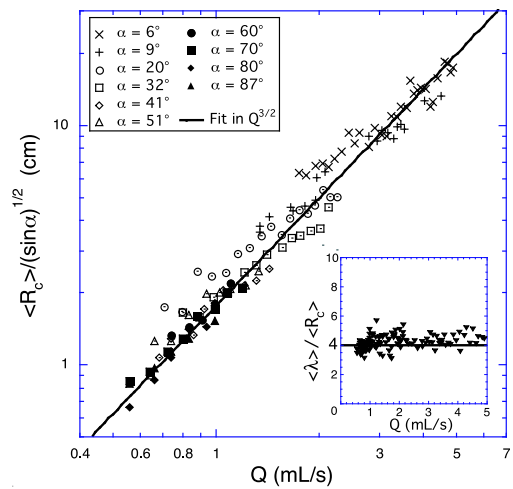}%
			\caption{Comparison of the data to the model for the mean radius of curvature. Insert: ratio of the wavelength to the radius of curvature for all plate inclinations $\alpha$.}%
		\label{Model}%
		\end{center}%
	\end{figure}

The magnitude of inertial and capillary forces depends on the shape of
the rivulet path. The equilibrium of gravity, inertia, capillarity and
pinning forces can therefore be used to solve for the expected radius
of curvature of stationary meanders. Again, gravity does not matter if
we are interested in the curvature at the vertical segments of the
meander. The force balance at the threshold of depinning/motion then
reads:
	\begin{equation}
		 \left( \rho Q^2/S - C(\theta)\gamma w \right)/R_{c} = \gamma \left( \cos\theta_{r}-\cos\theta_{a} \right)
	\end{equation}
This is the same as equation~(\ref{eq:instab}) except that $R_c$ is
now the final radius of curvature of the bends in the meander, which
scales between 1 and 6\,cm (see figure 3) and is much larger than the
initial radius of curvature $r_c$ considered for the threshold of the
instability. An order of magnitude calculation~\cite{forceestimates}
shows that the capillary contribution $C(\theta)\gamma w/R_c$ now
becomes smaller than pinning, and is also smaller than inertia $\rho
Q^2 /(S.R_c)$. We therefore neglect the capillary part and, assuming a
Poiseuille flow, one gets the following scaling law for the radius of
curvature:
	\begin{equation}
		\langle R_{c}\rangle \propto \frac{\rho (g/\nu)^{1/2}} { \gamma (\cos\theta_{r}-\cos\theta_{a})} Q^{3/2} \sqrt{\sin\alpha}
		\label{eq:R_c}
	\end{equation}
This expression has been tested on our data (Fig.~\ref{Model}) and the
scaling law fits the data reasonably well. The same scaling seems to
hold for amplitude and wavelength.

Some sets of data, for a given inclination, seem to have a slightly
different slope from the proposed scaling law. A first possible
explanation lies in the neglected capillary term, which would add a
contribution to the radius of curvature in
$-Q^{1/4}\sin^{-1/4}\alpha$, which is more important at low
inclinations and low flow rates, and diminishes the final radius. A
second important point is the reduction of the effective slope through
sinuosity \Si (ratio of total length of the rivulet to the distance
between its endpoints). The slope $\sin\alpha$ has to be replaced by
$\sin\alpha/\Si$ in equ~(\ref{eq:R_c}), and since \Si increases with
$Q$ and $\alpha$ \cite{Nakagawa84}, the exponents of $Q$ and $\alpha$
should be smaller than the ones in equ~\ref{eq:R_c}.
Last not least, the Poiseuille flow assumption might need to be
corrected. Indeed, flow visualizations with methylene blue dye show
regions with zero or even backward recirculating flow, already
mentioned by \cite{Walker}. The Reynolds number based on the rivulet
size is of order 500-1000.
\smallskip

{\it Discussion~--} The scenario of development of meanders may be
summed up as follows. Stable rivulets become unstable when inertia
dominates capillary and pinning forces. The destabilizing term
diminishes as a perturbation grows (and so its radius), so the
hysteresis of wetting eventually stops the growth of the
perturbation. In the final state, pinning forces should therefore be
maximally mobilized, with the critical advancing contact angle on the
outside of the curve and the critical receding contact angle on the
inside. Cross-section measurements by Nakagawa and
Scott~\cite{Nakagawa84} seem to corroborate this, but more
quantitative measurements are planned.

As shown in fig~\ref{Montage}\textit{b}, the final meanders are made
up of circular segments. The ratio $\langle\lambda\rangle/\langle R_c
\rangle$ is expected to be close to 4 for paths made of half-circles
possibly followed by horizontal segments. This is confirmed by the
insert in fig~\ref{Model} within experimental scatter.

In order to understand how the rivulet switches from one bend to the
next, gravity has to be taken into account to explain why the rivulet
does not follow the circular path back uphill. If weight is strong
enough, it will exceed pinning at a critical orientation of the
rivulet with respect to the slope, $\sin\Phi = \gamma(\cos\theta_r -
\cos\theta_a) / \rho g S \sin\alpha$ which will define the inflection
points on the path. If pinning always remains stronger, the rivulet
can become horizontal without slipping\,: horizontal segments appear,
in which the velocity decreases and the cross-section increases until
gravity becomes dominant. Both scenari seem to correspond to observed
patterns, but the effect on the amplitude of the meanders is not clear
yet.

A last point remains to be discussed\,: above $Q_{c1}$ perturbations
are stabilized at large scales because the destabilizing forces
decrease for smaller curvatures, but why do no new perturbations
appear at small scales? A possible reason might lie in the
considerable drop of the mean flow velocity (by more than a factor 2)
at the transition from straight-to-meandering, which has been revealed
by injection of methylene blue dye into the rivulets. The formation of
a meandering path decreases the mean slope seen by the rivulet, and
consequently the equilibrium velocity.  The rivulet therefore becomes
stable again in the meandering part, and remains unstable in the parts
which have not yet developed bends. It would be interesting to perform
velocity measurements to see whether the second threshold $Q_{c2}$ is
defined by the same critical local velocity as $Q_{c1}$.

Concerning the influence of viscosity on meanders, Schmuki and Laso
\cite{Schmuki} showed that viscosity damped meanders. Our scaling laws
are in agreement with this observation\,: for higher viscosity, both
the critical flow rate for the onset of meandering and the radius of
curvature increase. Other experiments, varying viscosity, should be
carried out to quantitatively determine the influence of viscosity on
meandering.

{\it Acknowledgments~--} We thank B. Andreotti and T. Podgorski for
stimulating discussions and critical reading of the manuscript.
\vspace{-2\baselineskip}



\begin{thebibliography}{}
\bibitem{linetension} Two forces act normal on a meander section $S$\,: pressure $-S \Delta p$ and surface tension. The surface tension is made up of two parts\,: the water-air interface of length $\ell$ contributes a term $\gamma\ell$, the straight mylar-water interface of length $w$ contributes a term $-\gamma w\cos\theta$ corresponding to the energy difference between wet and dry Mylar, $\theta$ being an average contact angle. The effective rivulet line tension is $T_{\gamma}=\gamma(\ell-w\cos\theta)-S\Delta p$.  Approximating the water-air interface by a circular arc of radius $r$ intersecting the solid plate at the contact angle $\theta$, we find that $\ell=2\theta r$, $w=2r\sin\theta$, $\Delta p=\gamma/r$ (neglecting the longitudinal rivulet curvature) and $S = \theta r^2-(1/2)wr\cos\theta$. The line tension becomes $T_{\gamma}= (1/2) \gamma w (\theta/\sin\theta - \cos\theta)=C(\theta) \gamma \omega$.
\bibitem{forceestimates} For the contact angles given in the text, the pinning force is around $F_h = 35\,\mathrm{g}.\mathrm{s}^{-2}$. Inertial and capillary terms evaluate to $\rho S v^2 = \rho Q^2 / S = 50\,\mathrm{g}.\mathrm{cm}.\mathrm{s}^{-1}$ and $C\gamma w = 6\,\mathrm{g}.\mathrm{cm}.\mathrm{s}^{-1}$ for $Q=1\,\mathrm{mL}.\mathrm{s}^{-1}$, $w=4\,\mathrm{mm}$.
\end{thebibliography}
\end{document}